# Multicommodity Flow Algorithms for Buffered Global Routing

Christoph Albrecht, Andrew B. Kahng,[*] Ion Măndoiu,[†] and Alexander Zelikovsky[‡]

Cadence Berkeley Labs, Berkeley, CA 94704, E-mail: calb@cadence.com

[*]CSE and ECS Departments, UC San Diego, La Jolla, CA 92093, E-mail: abk@ucsd.edu

[‡]CSE Department, University of Connecticut, Storrs, CT 06269, E-mail: ion@engr.uconn.edu

[‡]CS Department, Georgia State University, Atlanta, GA 30303, E-mail: alexz@cs.gsu.edu

## Abstract

In this paper we describe a new algorithm for buffered global routing according to a prescribed buffer site map. Specifically, we describe a provably good multi-commodity flow based algorithm that finds a global routing minimizing buffer and wire congestion subject to given constraints on routing area (wirelength and number of buffers) and sink delays. Our algorithm allows computing the tradeoff curve between routing area and wire/buffer congestion under any combination of delay and capacity constraints, and simultaneously performs buffer/wire sizing, as well as layer and pin assignment. Experimental results show that near-optimal results are obtained with a practical runtime.

## 1 Introduction

Due to delay scaling effects in deep-submicron technologies, interconnect planning and synthesis are becoming critical to meeting chip performance targets with reduced design turnaround time. In particular, the global routing phase of the design cycle is receiving renewed interest, as it must efficiently handle increasingly more complex constraints for increasingly larger designs (see [13] for a recent survey). In addition to handling traditional objectives such as congestion, wirelength and timing, a critical requirement for next generations of global routers is the integration with other interconnect optimizations, most importantly with buffer insertion and sizing. Indeed, it is estimated that top-level on-chip interconnect will require up to $10^6$ repeaters when we reach the 50nm technology node. Since these repeaters are large and have a significant impact on global routing congestion, buffer insertion and sizing can no longer be done after global routing completes.

In this paper, we present and enhance a powerful integrated approach introduced in [3] for congestion and timing-driven global routing, buffer insertion, pin assignment, and buffer/wire sizing. Our approach is based on a multicommodity flow formulation for the buffered global routing problem. Multicommodity flow based global routing has been an active research area since the seminal work of Raghavan and Thomson [16]. Although the global routing problem is NP-hard (even highly restricted versions of it, see [19]), [16] has shown that the optimum solution can be approximated arbitrarily close in time polynomial in the number of nets and the inverse of the accuracy. To date, predictability of solution quality continues to be a distinct advantage of multicommodity flow based methods over all other approaches to global routing, including popular rip-up-and-reroute approaches [13].

The original method of Raghavan and Thomson relies on randomized rounding of an *optimum* fractional multicommodity flow. Subsequent works [7, 14] have improved runtime scalability by using the approximation algorithm for multicommodity flows by [17]. Yet, only the recent breakthrough improvements due to Garg and Könemann [12] and Fleischer [11] have rendered multicommodity flow based global routing practical for full chip designs [2]. As [2], our algorithm is built upon the efficient multicommodity flow approximation scheme of [12, 11]. Our main contribution is a provably good multi-commodity flow based algorithm that, for a given buffer site map, finds a buffered global routing minimizing buffer and wire congestion subject to given constraints on routing area (wirelength and number of buffers) and sink delays.

The key features of our approach include the following:

- Our implementation permits detailed floorplan evaluation in that it enables computing the tradeoff curve between routing area and wire/buffer congestion under any combination of delay and capacity constraints.

- Like the allocation heuristic in [4], our algorithm enforces maximum source/buffer wireloads and controls congestion by taking into account routing channel capacities and buffer site locations. At the same time, like the buffer-block planning algorithm in [9], our algorithm takes into account individual sink delay constraints.

- Simultaneously, our algorithm performs buffer and wire sizing by taking into account given libraries of buffer types and wire widths, and integrates layer and pin assignment (the latter with virtually no increase in runtime). Soft pin locations are modeled as multiple sites (grid locations), and are enabling to solution quality.

The rest of the paper is organized as follows. In Section 2 we formalize the buffered global routing problem for 2-pin nets. Then, in Section 3.2, we reformulate the problem as a minimum cost integer multicommodity flow problem (with capacities on *sets of edges*), give an efficient algorithm for finding near-optimal solutions to the fractional relaxation, and show how to convert fractional solutions to near-optimal routings by randomized rounding. In Sections sec.multipin and 5 we show how our approach can be extended to handle multipin nets as well as pin assignment, polarity constraints imposed by the use of inverting buffers, buffer and wire sizing, and prescribed delay upperbounds. We conclude the paper with experimental results detailing the scalability and limitations of our algorithm and comparing it to the heuristic in [4].

## 2 Problem formulation

In this section we formulate the buffered global routing problem. To simplify the presentation, we ignore pin assignment flexibility and assume that there is a single (non-inverting) buffer type and a single wire width. We further assume that only buffer wireload constraints must be satisfied (i.e., ignore delay upper-bounds), and that each net has 2 pins only. Extensions of our approach to pin assignment, polarity constraints induced by the use of inverting buffers, buffer and wire sizing, timing constraints, and multi-pin nets are discussed in Section 5.

For a given floorplan and tile size, we construct a vertex- and edge-weighted *tile graph* $G = (V, E, b, w)$, $b: V \to \mathbb{N}$, $w: E \to \mathbb{N}$, where:

- $V$ is the set of tiles;

- $E$ contains an edge between any two adjacent tiles;

- For each tile $v \in V$, the *buffer capacity* $b(v)$ is the number of buffer sites located in $v$; and

- For each edge $e = (u, v) \in E$, the *wire capacity* $w(e)$ is the number of routing channels available between tiles $u$ and $v$.

We denote by $\mathcal{N} = \{N_1, N_2, \ldots, N_k\}$ the given netlist, where each net $N_i$ is specified by a *source* $s_i$ and a *sink* $t_i$.

A feasible solution to the buffered global routing problem seeks for each net $N_i$ an $s_i$–$t_i$ path $P_i$ buffered using the available buffer sites (see Figure 1.1) such that the source vertex and the buffers drive each at most $U$ units of wire, where $U$ is a given upper-bound (the example in Figure 1.1 has $U = 5$). Formally, a *feasible buffered routing* for net $N_i$ is a path $P_i = (v_0, v_1, \ldots, v_{l_i})$ in $G$ together with a set of buffers $B_i \subseteq \{v_0, \ldots, v_{l_i}\}$ such that:

- $v_0 = s_i$ and $v_{l_i} = t_i$;
- $w(v_{i-1}, v_i) \geq 1$ for every $i = 1, \ldots, l_i$;
- $b(v_i) \geq 1$ for every $v_i \in B_i$; and
- The length along $P_i$ between $v_0$ and the first buffer in $B_i$, between consecutive buffers, and between the last buffer and $v_{l_i}$, are all at most $U$.

We will denote by $\mathcal{R}_i$ the set of all feasible routings $(P_i, B_i)$ for net $N_i$. Given buffered routings $(P_i, B_i) \in \mathcal{R}_i$ for each net $N_i$, the relative *buffer congestion* is

$$\mu = \max_{v \in V} \frac{|\{i : v \in B_i\}|}{b(v)}$$

and the relative *wire congestion* is

$$\nu = \max_{e \in E} \frac{|\{i : e \in P_i\}|}{w(e)}$$

The buffered paths $(P_i, B_i)$, $i = 1, \ldots, k$, are simultaneously routable iff both $\mu \leq 1$ and $\nu \leq 1$. To leave resources available for subsequent optimization of critical nets and ECO routing, we will generally seek simultaneous buffered routings with buffer and wire congestion bounded away from 1. Using the total wire and buffer area as solution quality measure we get:

**Integrated Global Routing and Bounded Wireload Buffer Insertion Problem**[1]
**Given:**

- Grid-graph $G = (V, E, b, w)$, with buffer and wire capacities $b : V \to \mathbb{N}$, respectively $w : E \to \mathbb{N}$;
- Set $\mathcal{N} = \{N_1, \ldots, N_k\}$ of 2-pin nets with unassigned source and sink pins $S_i, T_i \subseteq V$; and
- Wireload, buffer congestion, and wire congestion upper-bounds $U > 0$, $\mu_0 \leq 1$, and $\nu_0 \leq 1$.

**Find:** feasible buffered routings $(P_i, B_i) \in \mathcal{R}_i$ for each net $N_i$ with relative buffer congestion $\mu \leq \mu_0$ and relative wire congestion $\nu \leq \nu_0$, minimizing the total wire and buffer area, i.e., $\alpha \sum_{i=1}^{k} |B_i| + \beta \sum_{i=1}^{k} |P_i|$, where $\alpha, \beta \geq 0$ are given constants.

## 3  Buffered Global Routing Via Multicommodity Flow Approximation

The high-level steps of our approach are the following:

1. Following Alpert et al. [4], we construct a 2-dimensional tile graph to capture the number and spatial distribution of wire routing tracks and buffer insertion sites available in the given floorplan. For simplicity, we assume throughout the paper that all tiles have the same size. As discussed in Section 6, uneven tiling, i.e, using fine tiling in highly congested regions of the design and coarse tiling in regions blocked for routing or buffering, can be used to improve the tradeoff between accuracy and solution time.

---
[1] The problem is called *Floorplan Evaluation Problem* in [3], but the formulation is useful in post-placement scenarios as well.

2. We then build an auxiliary graph in which every directed path from a net source to the net's sink captures a feasible wire route between them together with locations for the buffers to be inserted on this route such that buffer load constraints are satisfied. The auxiliary graph is obtained automatically from the tile graph using an original *gadget* construction (Section 3.1) which only increases the size of the graph by a linear factor.

3. We use the auxiliary graph to formulate the buffered global routing problem as an integer linear program (ILP). To formally express the ILP, we use a 0/1 variable for each source-sink path, and require that exactly one path be chosen for each source-sink pair. The objective is to minimize the wire and buffer congestion subject to a given upper-bound on the total wirelength (Section 3.1).

4. We find a near-optimal solution to the fractional relaxation of the above integer program. Although the integer program has exponential size (there are exponentially many variables corresponding to source-sink paths in the auxiliary graph), we give a combinatorial algorithm which runs in polynomial time by representing explicitly only non-zero variables (Section 3.2). The algorithm combines the general framework for multicommodity flow approximation introduced by [12, 11] with some of the extensions described in [2] and [10].

5. Finally, we round the fractional solution to an integer one using a heuristically enhanced version of the randomized rounding method originally proposed in [16] (Section 3.3).

In this section we detail each step of our approach for the case of 2-pin nets, and also discuss efficient computation of the entire tradeoff curve between routing area and congestion for a given floorplan (Section 3.4).

## 3.1 Gadget Graph and Integer Program Formulation

Recall that, for every feasible buffered routing in the tile graph $G = (V(G), E(G), b, w)$, the wireload of the source and of each buffer must be at most $U$. We start by defining an auxiliary directed graph $H$ which captures exactly these feasible buffered routings (see Figure 1.2). The graph $H$ has $U+1$ vertices $v^0, v^1, \ldots, v^U$ for each vertex $v \in V(G)$. The index of each copy corresponds to the *remaining wireload budget*, i.e., the number of units of wire that can still be driven by the last inserted buffer (or by the net's source). Buffer insertions are represented in the gadget graph by directed arcs of the form $(v^j, v^U)$ — following such an arc resets the remaining wireload budget up to the maximum value of $U$. Each undirected edge $(u, v)$ in the tile graph gives rise to directed arcs $(u^j, v^{j-1})$ and $(v^j, u^{j-1})$, $j = 1, \ldots, U$, in the gadget graph. Notice that the copy number decreases by 1 for each of these arcs, corresponding to a decrease of 1 unit in the remaining wireload budget. In addition, we add to $H$ individual vertices to represent net sources and sinks. Each source vertex is connected by a directed arc to the $U$-th copy of the node representing the enclosing tile. Furthermore, *all* copies of the nodes representing enclosing tiles are connected by directed arcs into the respective sink vertices.

Formally, the graph $H$ has vertex set

$$V(H) = \{s_i, t_i \,|\, 1 \leq i \leq k\} \cup \{v^j \,|\, v \in V(G), 0 \leq j \leq U\}$$

and arc set

$$E(H) = E_{src} \cup E_{sink} \cup \Big(\bigcup_{(u,v) \in E(G)} E_{u,v}\Big) \cup \Big(\bigcup_{v \in V(G)} E_v\Big)$$

where

$$\begin{aligned}
E_{src} &= \{(s_i, v^U) \,|\, \text{tile } v \text{ contains } s_i, 1 \leq i \leq k\} \\
E_{sink} &= \{(v^j, t_i) \,|\, \text{tile } v \text{ contains } t_i, 0 \leq j \leq U, 1 \leq i \leq k\} \\
E_{u,v} &= \{(u^j, v^{j-1}), (v^j, u^{j-1}) \,|\, 1 \leq j \leq U\} \\
E_v &= \{(v^j, v^U) \,|\, 0 \leq j < U\}
\end{aligned}$$

Each directed path in the gadget graph $H$ corresponds to a buffered routing in the tile graph, obtained by ignoring copy indices for tile vertices and replacing each "buffer" arc $(v^j, v^U)$ with a buffer inserted in tile $v$. Clearly, the construction ensures that the wireload of each buffer is at most $U$ since a directed path in $H$ can visit at most $U$ vertices before following a buffer arc. Therefore, we get:

**Lemma 1.1** *There is a 1-to-1 correspondence between the feasible buffered routings for net $N_i$ in the tile graph $G$ and the $s_i$–$t_i$ paths in $H$.*

We will use the correspondence established in Lemma 1.1 to give an integer linear program (ILP) formulation for the buffered global routing problem. Let $\mathcal{P}_i$ denote the set of all simple $s_i$–$t_i$ paths in $H$. We introduce a 0/1 variable $x_p$ for every path $p \in \mathcal{P} := \cup_1^k \mathcal{P}_i$. The variable $x_p$ is set to 1 if the buffered routing corresponding to $p \in P_i$ is used to connect net $N_i$, and to 0 otherwise. With this notation, the buffered global routing problem can be formulated as follows:

$$\min \sum_{p \in \mathcal{P}} \left( \alpha \sum_{v \in V(G)} |p \cap E_v| + \beta \sum_{(u,v) \in E(G)} |p \cap E_{u,v}| \right) x_p \qquad (1.1)$$

subject to

$$\sum_{p \in \mathcal{P}} |p \cap E_v| x_p \leq \mu_0 \, b(v), \qquad v \in V(G)$$

$$\sum_{p \in \mathcal{P}} |p \cap E_{u,v}| x_p \leq \nu_0 \, w(u,v), \qquad (u,v) \in E(G)$$

$$\sum_{p \in \mathcal{P}_i} x_p = 1, \qquad i = 1, \ldots, k$$

$$x_p \in \{0, 1\}, \qquad p \in \mathcal{P}$$

ILP (1.1) is similar to the "path" formulation of the classical minimum cost integer multicommodity flow problem [1]. The only difference is that capacity constraints on the edges and vertices of the tile graph $G$ become capacity constraints for sets of edges of the gadget graph $H$ (see Figure 1.2). We note that the buffered global routing problem can be represented more compactly by using a polynomial number of edge-flow variables instead of the exponential number of path-flow variables $x_p$. However, we use formulation (1.1) since it is the natural setting for describing the approximation algorithm in next section. The exponential number of variables is not impeding the efficiency of the approximation algorithm, which, during its execution, represents explicitly only a polynomial number of paths with non-zero flow.

## 3.2 The approximation algorithm

In this section we give an efficient approximation algorithm that can be used for solving the fractional relaxation of ILP (1.1). Using an approach similar to that used in [12] for solving the minimum cost concurrent multicommodity flow problem (see also [2]), instead of solving the relaxation of ILP (1.1) directly we introduce an upper bound $D$ on the wire and buffer area and consider the following linear program (LP):

$$\min \lambda \qquad (1.2)$$

subject to

$$\sum_{p \in \mathcal{P}} \left( \alpha \sum_{v \in V(G)} |p \cap E_v| + \beta \sum_{(u,v) \in E(G)} |p \cap E_{u,v}| \right) x_p \leq \lambda \, D$$

$$\sum_{p \in \mathcal{P}} |p \cap E_v| x_p \leq \lambda \, \mu_0 \, b(v), \qquad v \in V(G)$$

$$\sum_{p \in \mathcal{P}} |p \cap E_{u,v}| x_p \leq \lambda \, \nu_0 \, w(u,v), \qquad (u,v) \in E(G)$$

$$\sum_{p \in \mathcal{P}_i} x_p = 1, \qquad i = 1, \ldots, k$$

$$x_p \geq 0, \qquad p \in \mathcal{P}$$

Let $\lambda^*$ be the optimum objective value for LP (1.2). Solving the fractional relaxation of ILP (1.1) is equivalent to finding the minimum $D$ for which $\lambda^* \leq 1$. This can be done by a binary search which requires solving the LP (1.2) for each probed value of $D$. A lower bound on the optimal value of $D$ can be derived by ignoring all buffer and wire capacity constraints, i.e., by computing for each net $N_i$ buffered paths $p \in \mathcal{P}_i$ minimizing $\alpha \sum_{v \in V(G)} |p \cap E_v| + \beta \sum_{(u,v) \in E(G)} |p \cap E_{u,v}|$. A trivial upper bound is the total routing area available, i.e., $D_{\max} = \alpha \mu_0 \sum_{v \in V(G)} b(v) + \beta \nu_0 \sum_{(u,v) \in E(G)} w(u,v)$. In particular, unfeasibility in the fractional relaxation of ILP (1.1) is equivalent to $\lambda^*$ being greater than 1 when $D = D_{\max}$, and can therefore be detected using the algorithm described below.

The algorithm for approximating the optimum solution to LP (1.2) (see Figure 1.3) uses the general framework for multicommodity flow approximation introduced in [12] combined with ideas similar to those in [10] for efficiently handling set capacity constraints. The algorithm relies on simultaneously approximating the *dual* LP:

$$\max \sum_{i=1}^{k} l_i \quad (1.3)$$

subject to

$$\sum_{v \in V(G)} \mu_0 b(v) y_v + \sum_{(u,v) \in E(G)} \nu_0 w(u,v) z_{u,v} + Du = 1$$

$$\sum_{v \in V(G)} |p \cap E_v|(y_v + \alpha u) + \sum_{(u,v) \in E(G)} |p \cap E_{u,v}|(z_{u,v} + \beta u) \geq l_i, \ p \in \mathcal{P}_i$$

$$y_v \geq 0, \quad v \in V(G)$$
$$z_e \geq 0, \quad e \in E(G)$$

The algorithm starts with trivial solutions for LPs (1.2) and (1.3), and then updates these solutions over several phases. In each phase (lines 5–16 in Figure 1.3) one unit of flow is routed for each commodity; a feasible solution to LP (1.2) is obtained in line 17 after dividing all path flows by the number of phases. Commodities are routed along paths with minimum weight with respect to weights of $y_v + \alpha u$ for arcs in $E_v$, $v \in V(G)$, of $z_{u,v} + \beta u$ for arcs in $E_{u,v}$, $(u,v) \in E(G)$, and of 0 for all the other arcs (cf. LP (1.3)). The dual variables are increased by a multiplicative factor for all vertices/edges on a routed path; this ensures that dual weights increase exponentially with usage and thus often used edges are subsequently avoided [12].

Minimum-weight paths are computed in line (11) of the algorithm using Dijkstra's single-source shortest path algorithm. To reduce the number of shortest path computations, paths are recomputed only when their weight increases by a factor of more than $(1 + \gamma\varepsilon)$ (see the test in line 9). This speed-up idea, first applied in [11] for the maximum multicommodity flow problem, has been shown in [2] to decrease the running time in practice while maintaining the same theoretical worst-case runtime.

**Theorem 1.1** *The algorithm in Figure 1.3 finds an $(1 + \varepsilon_0)$-approximation with $O\left(\frac{1}{\varepsilon_0^2 \lambda^*} k \log n\right)$ minimum-weight path computations, using $\varepsilon = \min\left\{\frac{1}{\gamma}, \frac{1}{\gamma}(\sqrt{1+\varepsilon_0}-1), \frac{1}{4}\left(1-\left(\frac{1}{1+\varepsilon_0}\right)^{1/6}\right)\right\}$ and $\delta = \left(\frac{1-\varepsilon'}{n+m}\right)^{1/\varepsilon}$, where $n$ and $m$ are the number of vertices and edges of $G$, and $\varepsilon' := \varepsilon(1+\varepsilon)(1+\varepsilon\gamma)$.*

*Proof:* Let $t$ be the total number of phases executed by the algorithm. We will prove that if the algorithm had stopped one phase before the last one, namely after $t-1$ phases, the solution would have had the desired approximation ratio.

Let $y_v^{(r,i)}$, $z_{u,v}^{(r,i)}$ and $u^{(r,i)}$ be the value of variables $y_v$, $z_{u,v}$ and $u$ after net $i$ has been considered in phase $r$ and the variables have been increased in line 14, and let $y_v^{(r)} := y_v^{(r,k)}$, $z_{u,v}^{(r)} := z_{u,v}^{(r,k)}$ and $u^{(r)} := u^{(r,k)}$ be the value at the end of phase $r$. At the beginning we have

$$\mu_0 \sum_{v \in V(G)} b(v) y_v^{(0)} + \nu_0 \sum_{(u,v) \in E(G)} w(u,v) z_{u,v}^{(0)} + Du^{(0)} = \mu_0 \sum_{v \in V(G)} b(v) \frac{\delta}{\mu_0 b(v)} + \nu_0 \sum_{(u,v) \in E(G)} w(u,v) \frac{\delta}{\nu_0 w(u,v)} + D\frac{\delta}{D} = (n+m+1)\delta \quad (1.4)$$

When the dual variables $y_v$, $z_{u,v}$ and $u$ are increased in line 13 after a path $p_i \in \mathcal{P}_i$ has been found, we have

$$\mu_0 \sum_{v \in V(G)} b(v) y_v^{(r,i)} + \nu_0 \sum_{(u,v) \in E(G)} w(u,v) z_{u,v}^{(r,i)} + D u^{(r,i)}$$

$$= \mu_0 \sum_{v \in V(G)} b(v) y_v^{(r,i-1)} \left(1 + \varepsilon \frac{|p_i \cap E_v|}{\mu_0 b(v)}\right) + \nu_0 \sum_{(u,v) \in E(G)} w(u,v) z_{u,v}^{(r,i-1)} \left(1 + \varepsilon \frac{|p_i \cap E_{u,v}|}{\nu_0 w(u,v)}\right)$$

$$+ D u^{(r,i-1)} \left(1 + \varepsilon \frac{\alpha \sum_{v \in V(G)} |p_i \cap E_v| + \beta \sum_{(u,v) \in E(G)} |p_i \cap E_{u,v}|}{D}\right)$$

$$= \mu_0 \sum_{v \in V(G)} b(v) y_v^{(r,i-1)} + \nu_0 \sum_{(u,v) \in E(G)} w(u,v) z_{u,v}^{(r,i-1)} + D u^{(r,i-1)}$$

$$+ \varepsilon \left( \sum_{v \in V(G)} |p_i \cap E_v| (y_v^{(r,i-1)} + \alpha u^{(r,i-1)}) + \sum_{(u,v) \in E(G)} |p_i \cap E_{u,v}| (z_{u,v}^{(r,i-1)} + \beta u^{(r,i-1)}) \right)$$

In the following phases a path for net $i$ is only computed if the cost of the last path found has increased by more than $(1 + \gamma \varepsilon)$. Since the dual variables are increased only during the algorithm, we always have at the beginning of line 13:

$$\sum_{v \in V(G)} |p_i \cap E_v| (y_v^{(r,i-1)} + \alpha u^{(r,i-1)}) + \sum_{(u,v) \in E(G)} |p_i \cap E_{u,v}| (z_{u,v}^{(r,i-1)} + \beta u^{(r,i-1)})$$

$$\leq (1 + \gamma \varepsilon) l_i$$

$$\leq (1 + \gamma \varepsilon) \left( \min_{p \in \mathcal{P}_i} \sum_{v \in V(G)} |p \cap E_v| (y_v^{(r)} + \alpha u^{(r)}) + \sum_{(u,v) \in E(G)} |p \cap E_{u,v}| (z_{u,v}^{(r)} + \beta u^{(r)}) \right).$$

which means that at the end of phase $r$ we have:

$$\mu_0 \sum_{v \in V(G)} b(v) y_v^{(r)} + \nu_0 \sum_{(u,v) \in E(G)} w(u,v) z_{u,v}^{(r)} + D u^{(r)}$$

$$\leq \mu_0 \sum_{v \in V(G)} b(v) y_v^{(r-1)} + \nu_0 \sum_{(u,v) \in E(G)} w(u,v) z_{u,v}^{(r-1)} + D u^{(r-1)}$$

$$+ \varepsilon (1 + \gamma \varepsilon) \sum_{i=1}^{k} \min_{p \in \mathcal{P}_i} \left( \sum_{v \in V(G)} |p_i \cap E_v| (y_v^{(r)} + \alpha u^{(r)}) + \sum_{(u,v) \in E(G)} |p_i \cap E_{u,v}| (z_{u,v}^{(r)} + \beta u^{(r)}) \right) \quad (1.5)$$

For brevity, we set $\varepsilon' := \varepsilon(1 + \gamma \varepsilon)$. By linear programming duality, the expression

$$\lambda_{lb}^{(r)} := \frac{\sum_{i=1}^{k} \min_{p \in \mathcal{P}_i} \left( \sum_{v \in V(G)} |p_i \cap E_v| (y_v^{(r)} + \alpha u^{(r)}) + \sum_{(u,v) \in E(G)} |p_i \cap E_{u,v}| (z_{u,v}^{(r)} + \beta u^{(r)}) \right)}{\mu_0 \sum_{v \in V(G)} b(v) y_v^{(r)} + \nu_0 \sum_{(u,v) \in E(G)} w(u,v) z_{u,v}^{(r)} + D u^{(r)}}$$

is a lower bound on the maximum relative congestion, that is $\lambda_{lb}^{(r)} \leq \lambda^*$.

With this, inequality (1.5) can be rewritten as

$$\mu_0 \sum_{v \in V(G)} b(v) y_v^{(r)} + \nu_0 \sum_{(u,v) \in E(G)} w(u,v) z_{u,v}^{(r)} + D u^{(r)}$$

$$\leq \mu_0 \sum_{v \in V(G)} b(v) y_v^{(r-1)} + \nu_0 \sum_{(u,v) \in E(G)} w(u,v) z_{u,v}^{(r-1)} + D u^{(r-1)}$$

$$+ \varepsilon' \lambda_{lb}^{(p)} \left( \mu_0 \sum_{v \in V(G)} b(v) y_v^{(r)} + \nu_0 \sum_{(u,v) \in E(G)} w(u,v) z_{u,v}^{(r)} + D u^{(r)} \right)$$

which can be transformed to

$$\mu_0 \sum_{v \in V(G)} b(v) y_v^{(r)} + \nu_0 \sum_{(u,v) \in E(G)} w(u,v) z_{u,v}^{(r)} + D u^{(r)}$$
$$\leq \frac{1}{1 - \varepsilon' \lambda_{lb}^{(p)}} \left( \mu_0 \sum_{v \in V(G)} b(v) y_v^{(r-1)} + \nu_0 \sum_{(u,v) \in E(G)} w(u,v) z_{u,v}^{(r-1)} + D u^{(r-1)} \right)$$

If we set $\lambda_{lb} := \max_{r=1,\ldots,t} \lambda_{lb}^{(r)}$, we get with equation (1.4):

$$\begin{aligned}
\mu_0 \sum_{v \in V(G)} b(v) y_v^{(r)} + \nu_0 \sum_{(u,v) \in E(G)} w(u,v) z_{u,v}^{(r)} + D u^{(r)} &\leq \frac{(n+m+1)\delta}{(1 - \varepsilon' \lambda_{lb})^r} \\
&= \frac{(n+m+1)\delta}{(1 - \varepsilon' \lambda_{lb})} \left( 1 + \frac{\varepsilon' \lambda_{lb}}{1 - \varepsilon' \lambda_{lb}} \right)^{r-1} \\
&\leq \frac{(n+m+1)\delta}{(1 - \varepsilon')} \left( 1 + \frac{\varepsilon' \lambda_{lb}}{1 - \varepsilon'} \right)^{r-1} \\
&\leq \frac{(n+m+1)\delta}{(1 - \varepsilon')} e^{\frac{\varepsilon' \lambda_{lb}(r-1)}{1 - \varepsilon'}}
\end{aligned} \quad (1.6)$$

For the last inequality, $1 + x \leq e^x$ for $x \geq 0$ is used.

An upper bound on $\lambda$ for the solution computed by the algorithm can now be derived: We will demonstrate this for the second constraint in (1.2), for the first and third constraint the same upper bound can be obtained similarly.

Suppose node $v$ is used $s$ times by some path during the first $t-1$ phases, and let the $j$th increment, by which the constraint $\frac{1}{\mu_0 b(v)} \sum_{p \in \mathcal{P}_i} |p \cap E_v| x_p \leq \lambda$ is incremented, be $a_j := \frac{1}{\mu_0 b(v)} |p_i \cap E_v|$ for the appropriate $i$. After rescaling the variables $x_p, p \in \mathcal{P}$, we have $\frac{1}{\mu_0 b(v)} \sum_{p \in \mathcal{P}_i} |p \cap E_v| x_p = \sum_{j=1}^{s} a_j/(t-1)$. Since $y_v^{(0)} = \frac{\delta}{\mu_0 b(v)}$ and $y_e^{(t-1)} < \frac{1}{\mu_0 b(v)}$ (because the condition in line 4 still holds before the last phase is executed) and since

$$y_v^{(t-1)} = \frac{\delta}{\mu_0 b(v)} \prod_{j=1}^{s} (1 + \varepsilon a_j),$$

we get

$$\frac{\delta}{\mu_0 b(v)} \prod_{j=1}^{s} (1 + \varepsilon a_j) < \frac{1}{\mu_0 b(v)}.$$

Since $(1 + \varepsilon)^a \leq 1 + \varepsilon a$ for $0 \leq a \leq 1$ (the expression $(1 + \varepsilon)^a$ is convex in $a$, $1 + \varepsilon a$ is linear and we have equality for $a = 0$ and $a = 1$), it follows

$$(1 + \varepsilon)^{\sum_{j=1}^{s} a_j} < \frac{1}{\delta},$$

and hence we get

$$\frac{1}{\mu_0 b(v)} \sum_{p \in \mathcal{P}_i} |p \cap E_v| x_p = \frac{\sum_{j=1}^{s} a_j}{t-1} \leq \frac{\log_{1+\varepsilon} \left( \frac{1}{\delta} \right)}{t-1}.$$

As we can derive the same bounds for the first and second constraint, we have

$$\lambda^{(t-1)} \leq \frac{\log_{1+\varepsilon} \left( \frac{1}{\delta} \right)}{t-1}. \quad (1.7)$$

Since $\mu_0 \sum_{v \in V(G)} b(v) y_v^{(r)} + \nu_0 \sum_{(u,v) \in E(G)} w(u,v) z_{u,v}^{(r)} + D u^{(r)} \geq 1$, solving inequality (1.6) with $r = t$ for $\lambda_{lb}$ gives a lower bound on the optimum solution value:

$$\lambda_{lb} \geq \frac{1 - \varepsilon'}{\varepsilon'(t-1)} \ln \left( \frac{1 - \varepsilon'}{(n+m+1)\delta} \right),$$

from which together with (1.7) we get an upper bound on the approximation ratio $\rho$:

$$\rho \leq \frac{\lambda^{(t-1)}}{\lambda_{lb}} \leq \frac{\frac{\log_{1+\varepsilon}\left(\frac{1}{\delta}\right)}{t-1}}{\frac{1-\varepsilon'}{\varepsilon'(t-1)} \ln\left(\frac{1-\varepsilon'}{(n+m+1)\delta}\right)}$$

$$= \frac{\varepsilon'}{(1-\varepsilon')\ln(1+\varepsilon)} \frac{\ln\left(\frac{1}{\delta}\right)}{\ln\left(\frac{1-\varepsilon'}{(n+m+1)\delta}\right)}.$$

If $\delta$ is now chosen to be $\delta := \left(\frac{1-\varepsilon'}{(n+m+1)\delta}\right)^{\frac{1}{\varepsilon'}}$, we get

$$\frac{\ln(\frac{1}{\delta})}{\ln\left(\frac{1-\varepsilon'}{(n+m+1)\delta}\right)} = \frac{1}{1-\varepsilon'}$$

such that

$$\rho \leq \frac{\varepsilon'}{(1-\varepsilon')^2 \ln(1+\varepsilon)}$$

$$\leq \frac{\varepsilon(1+\gamma\varepsilon)}{(1-\varepsilon(1+\gamma\varepsilon))^2 \left(\varepsilon - \frac{\varepsilon^2}{2}\right)}$$

$$= \frac{1+\gamma\varepsilon}{(1-\varepsilon(1+\gamma\varepsilon))^2 \left(1 - \frac{\varepsilon}{2}\right)}$$

If $\varepsilon \leq \frac{1}{\gamma}$, we have $1 + \gamma\varepsilon \leq 2$ and we get

$$\rho \leq \frac{1+\gamma\varepsilon}{(1-2\varepsilon)^3}$$

If $\varepsilon$ is chosen such that $1 + \gamma\varepsilon \leq \sqrt{1+\varepsilon_0}$ and $\frac{1}{(1-2\varepsilon)^3} \leq \sqrt{1+\varepsilon_0}$, so we choose

$$\varepsilon = \min\left(\frac{1}{\gamma}, \frac{1}{\gamma}\left(\sqrt{1+\varepsilon_0} - 1\right), \frac{1}{2}\left(1 - \left(\frac{1}{1+\varepsilon_0}\right)^{\frac{1}{6}}\right)\right),$$

we get $\rho \leq 1 + \varepsilon_0$. After all, we have $\varepsilon = O(\varepsilon_0)$ and also $\varepsilon' = O(\varepsilon_0)$.

From (1.7) we get that

$$\lambda^* \leq \frac{\log_{1+\varepsilon}\left(\frac{1}{\delta}\right)}{t-1},$$

which means that the maximum number of phases is bounded by:

$$t \leq 1 + \frac{\log_{1+\varepsilon}\left(\frac{1}{\delta}\right)}{\lambda^*}$$

$$= 1 + \frac{1}{\ln(1+\varepsilon)\varepsilon'\lambda^*} \ln\left(\frac{n+m+1}{1-\varepsilon'}\right).$$

The last expression can be bounded by $O\left(\frac{1}{\varepsilon_0^2 \lambda^*} \ln n\right)$. □

**Remark 1.** The dependence on $\lambda^*$ in Theorem 1.1 can be eliminated by a scaling technique described in [11]. Thus, using a Fibonacci heap implementation of Dijkstra's algorithm to compute minimum-weight paths leads to a runtime of $O\left(\frac{1}{\varepsilon_0^2} k(m + n \log n) \log n\right)$ for the algorithm in Figure 1.3.

**Remark 2.** Using ideas from [2], it can be shown that the algorithm in Figure 1.3 not only minimizes $\lambda$, but also "strives" for a lexicographically minimum solution with respect to the vector consisting of the relative buffer congestion of the vertices, the relative wire congestion of the edges, and the ratio between the total routing area and the upperbound $D$. This is particularly useful for the case when the floorplan is unroutable using the given buffer sites and wire tracks, since then the solution returned by the algorithm indicates where we should add more routing resources (by local perturbations of the floorplan) to reach routability. As noted above, this case is handled by running the algorithm with $D = D_{\max}$. If we want to completely ignore the constraint on total routing area (i.e., set $D = \infty$), the dual variable $u$ is 0 throughout the whole execution of the algorithm and can thus be eliminated.

**Remark 3.** In a practical implementation, line (2) of the algorithm, which requires setting to zero an exponential number of variables, is not implemented explicitly. Rather, the algorithm keeps track only of the paths with non-zero flow, i.e., those paths for which flow is augmented in line (13). Several alternatives for memory efficient representation of the non-zero flows are discussed in next section.

## 3.3 Randomized Rounding

In the previous section we have presented an algorithm for computing near-optimal solutions to LP 1.2. The last step in solving the buffered global routing problem is to convert these fractional flows into feasible buffered routings for each net. We follow the randomized rounding technique proposed by Raghavan and Thomson [16], and route each net $N_i$ by randomly choosing one of the paths $p \in \mathcal{P}_i$, where the probability of choosing path $p$ is equal to the fractional flow $x_p$ (recall that $\sum_{p \in \mathcal{P}_i} x_p = 1$, i.e., $x_p$ is a probability distribution over $\mathcal{P}_i$). Since the fractional flows satisfy buffer and wire congestion constraints, it follows that (for large enough capacities) the relative congestion after rounding increases only by a small factor [16].

A direct implementation of the randomized rounding scheme requires storing explicitly all paths with non-zero flow. However, this is typically unfeasible due to the large memory requirement. An alternative implementation, originally suggested by [16], is to compute edge flows instead of path flows during the algorithm in Figure 1.3, and then implement randomized rounding by performing a random walk between the source and sink of each net. As noted in [10], performing the random walks *backwards*, i.e., from sinks towards sources, leads to reduced congestion for the case when a significant number of the 2-pin nets result from decomposition of multi-pin nets.

A simpler implementation, which requires storing a single path per net, is to interleave randomized rounding with the computation of the fractional flows $x_p$. In this implementation, we continuously update the path selected for each net, as follows. In first phase, the single path routed for each net becomes the net's choice with probability 1. In iteration $r > 1$, the path routed for net $i$ replaces the previous selection of net $i$ with a probability of $(r-1)/r$. It is easy to see that the path selected after $t$ phases was selected by the net in phase $r = 1, \ldots, t$ with an equal probability of $1/r$, i.e., the probability that a path $p$ is the final selection is equal to the fractional flow $x_p$ computed by the algorithm in Figure 1.3.

The results reported in Section 7 were obtained using yet another implementation. In our implementation we save the paths routed for each net in the *last $K = 5$ phases* of the algorithm in Figure 1.3 (note that the $K$ paths resulting for each net need not be distinct). Then, we pick for each net one of the $K$ saved paths, uniformly at random. To further improve the results, we repeat the random choices a large number of times (10,000 times in our implementation) and keep the choices that result in the smallest congestion or routing area (depending on the optimization criteria). We found this scheme, which has still reasonable memory requirements, to work better in practice than the other approaches (although, technically, it implements only a rough approximation of the probability distribution required by [16]).

## 3.4 Area and congestion tradeoff curve

To evaluate a floorplan at an early stage of the design process, it is useful to find not only the minimum routing area needed for given bounds on $\mu_0$ and $\nu_0$ on the relative buffer and wire congestion, but also how the total routing area increases if we enforce a smaller congestion. Obviously, a floorplan is better if a smaller area increase is needed for the same decrease in congestion. Let us denote by $\Lambda(\mu, \nu)$ the minimum routing area needed for a fractional solution with relative buffer and wire congestion not more than $\mu$ and $\nu$, respectively. In the following, we denote a fractional solution $x_p$, $p \in \mathcal{P}$ for LP (1.2), simply by a vector $x$. Let $A(x)$, $\mu(x)$, and $\nu(x)$ denote the total routing area, buffer congestion, respectively wire congestion of $x$.

**Lemma 1.2** *The function $(\mu, \nu) \longmapsto \Lambda(\mu, \nu)$ is convex.*

*Proof:* Let $x$ be an optimal solution (minimal routing area) for given relative congestions $\mu$ and $\nu$, that is $A(x) = \Lambda(\mu, \nu)$, $\mu(x) \leq \mu$, $\nu(x) \leq \nu$, and, similarly, let $x'$ be an optimal solution for relative congestions $\mu'$ and $\nu'$. For any given $\kappa \in [0, 1]$, we will now construct a solution $x''$ with maximum relative buffer congestions at most $\mu'' = \kappa\mu + (1 - \kappa)\mu'$, wire congestion at most $\nu'' = \kappa\nu + (1 - \kappa)\nu'$, and area $A(x'') = \kappa A(x') + (1 - \kappa)A(x')$, and hence $\Lambda(\mu'', \nu'') \leq \kappa\Lambda(\mu, \nu) + (1 - \kappa)\Lambda(\mu', \nu')$.

The solution $x'' := \kappa x + (1 - \kappa)x'$ fulfills all these requirements. It describes a feasible routing: Each net $i$ is routed exactly once, because $\sum_{p \in \mathcal{P}_i} x''_p = \sum_{p \in \mathcal{P}_i} (\kappa x_p + (1 - \kappa)x'_p) = \kappa \sum_{p \in \mathcal{P}_i} x_p + (1 - \kappa) \sum_{p \in \mathcal{P}_i} x'_p = \kappa 1 + (1 - \kappa)1 = 1$, and we have $x''_p = \kappa x_p + (1 - \kappa)x'_p \geq 0$ for each $p \in \mathcal{P}$. For the relative congestion we get: $\mu(x'') = \max_{v \in V(G)} \frac{1}{b(v)} \sum_{p \in \mathcal{P}} |p \cap E_v| x''_p = \max_{v \in V(G)} \frac{1}{b(v)} \sum_{p \in \mathcal{P}} |p \cap E_v| (\kappa x_p + (1 - \kappa)x'_p) \leq \kappa \max_{v \in V(G)} \frac{1}{b(v)} \sum_{p \in \mathcal{P}} |p \cap E_v| x_p + (1 - \kappa) \max_{v \in V(G)} \frac{1}{b(v)} \sum_{p \in \mathcal{P}} |p \cap E_v| x'_p \leq \kappa\mu + (1 - \kappa)\mu' = \mu''$, and similarly $\nu(x'') \leq \nu''$. Since the area is a linear function in $x$, we get also $A(x'') = \kappa A(x) + (1 - \kappa)A(x')$, which concludes the proof. □

The following lemma shows that in certain cases we can derive a value $\Lambda(\mu, \nu)$ from an optimal solution of the linear program (1.2), and thus the binary search suggested in Section 3.2 can be avoided.

**Lemma 1.3** *Let $x$ be an optimal solution of LP (1.2) for given $D$, $\mu_0$ and $\nu_0$. If there exists a solution $x'$ with $\max\left(\frac{\mu(x')}{\mu_0}, \frac{\nu(x')}{\nu_0}\right) < \max\left(\frac{\mu(x)}{\mu_0}, \frac{\nu(x)}{\nu_0}\right)$, then $\Lambda(\mu(x), \nu(x)) = A(x)$*

*Proof:* Suppose there exists a solution $x''$ with $A(x'') < A(x)$ and $\mu(x'') \leq \mu(x)$, $\nu(x'') \leq \nu(x)$. Then the convex combination $x''' := x'' + \varepsilon(x' - x'')$, $\varepsilon > 0$, $\varepsilon$ small, gives a better solution for LP (1.2). □

This suggests the following approach to computing the full area vs. congestion tradeoff curve. First, compute the feasible region (which is also convex) for $\mu$ and $\nu$ by ignoring the constraint on the area. Then solve the linear program LP (1.2) for certain values of $D$, $\mu_0$ and $\nu_0$. If the solution is on the boundary of the feasible region, decrease $D$ such that $\mu$ and $\nu$ increase, otherwise a new point for the area and congestion tradeoff curve has been found.

## 4 Handling Multipin Nets

Although a majority of nets have only two pins, modern designs also contain an increasing number of multipin nets. In order to apply our approximation algorithm from the previous section, we need to reduce multipin nets to 2-pin nets or otherwise adjust the algorithm to handle mutipin nets. In this section we consider several methods for decomposing multipin nets and present an extension of our algorithm to 3-pin nets; these alternatives provide a range of trade-offs between runtime and solution quality.

The standard reduction constructs the minimum spanning tree $T$ over all $k$ pins of a $k$ pin net and then splits the net into $k - 1$ 2-pin nets each corresponding to a single edge in $T$. Although the wireload can be accurately taken in account, the inherent drawback of this approach is that for high fanout nets we may end up with unbalanced and overloaded buffering. Note that the star topology decomposition, in which the source is connected with each sink by a separate edge, suffers from the same drawback. Instead of spanning trees, we suggest to use buffered Steiner trees. The minimum buffered routing for the entire $k$-pin net can be found using one of the algorithms from [6]. Such routing has been shown to be very close to optimal and is convenient for handling high fanout nets – both sink and wire loads are taken into account. The resulting buffered routing tree $T$ connects the vertices of degree 1 (corresponding to terminals), degree 2 (corresponding to buffers) and degree 3 and 4 (corresponding to branching points for Steiner tree routing) (see Figure 1.4(a)). Our approach is to split $T$ into smaller pieces (2- or 3-pin nets) and route them separately using the algorithm from the previous section. The resulting pieces have longer total wirelength, thus allowing more flexibility for congestion minimization. We distinguish three methods of splitting $T$.

**Fixed branching and fixed buffering.** The tree $T$ is split into 2-vertex subgraphs by replacing each vertex $v$ with the $deg(v)$ copies each corresponding to one of the incident edges (see Figure 1.4(b)). Each subgraph corresponds to a single edge of $T$ and has a single source and a single sink. The number of resulting 2-pin nets is $k + p + b - 1$, where $k$ is the number of terminals, $p$ is the number of Steiner points and $b$ is the number of buffers in $T$. This decomposition is the least flexible - the positions of both buffers and Steiner points are fixed. Routing- and buffer-congestion minimization may require using very long detours with possible addition of extra buffers.

**Fixed branching and flexible buffering.** The tree $T$ is split into 2-vertex subgraphs by replacing each branching vertex $v$ with the $deg(v)$ copies, each corresponding to one of the incident edges (see Figure 1.4(c)). Each subgraph corresponds to a single edge of the unbuffered version of $T$. The number of resulting 2-pin nets is $k+p-1$. Similar to the previous decomposition, there is limited room for rerouting but now the buffers can be shifted between grid cells resulting in much better opportunities for buffer-congestion minimization. Since the buffer insertion may be caused by large sink load (e.g., in Figure 1.4(c) net $(s, p_1)$ requires a buffer because of the 3 sinks downstream), it is necessary to compensate the upper bound $U$ for some nets. This can be easily done by decreasing the level of the source of such net as follows. Normally, the source of a net is placed at the highest node of the edge gadget, e.g., $u^5$ on Figure 1.2, but if compensation is necessary, then the source should be placed lower, e.g., $u^1$, thus requiring a buffer much sooner.

**Half-flexible branching and flexible buffering.** Assume for simplicity that all Steiner points in $T$ have degree 3 i.e., no degree-4 Steiner points are allowed. The vertices of the unbuffered tree $T$ can be colored into 2 colors such that adjacent vertices have different color. If we fix locations of the Steiner points of the same color then all the resulting nets will have either 3 or 2 terminals (see Figure 1.4(d)). If we pick the color with the least amount of Steiner points, then we can be sure that at most half of all Steiner points are fixed implying that the number of resulting nets is at most $k - 1$. Indeed, at least $p/2$ Steiner points are not fixed and the corresponding three 2-pin nets for fixed branching are replaced with one 3-pin net. This reduces the total number of nets with respect to fixed branching by at least $2p/2 = p$.

This decomposition is the most flexible one, giving most opportunities for routing and buffer-congestion minimization. Unfortunately, it requires runtime-costly adjustments to the approximation algorithm from the previous section. Figure 1.5 gives the modified subroutine for computing feasible routings having minimum weight with respect to the dual variables. We assume here that the possible locations of the source pin for a net $N_i$ are specified by $S_i$ as before, while the two sinks are specified by sets $T_i^1$ and $T_i^2$. In the graph $H$ we have vertices $t_i^1$ and $t_i^2$ and edge sets $\{(v^j, t_i^l) \mid v \in T_i^l, j = 0, ..., U\}$, $l = 1, 2$ for the sink pins of such a 3-pin net. For each possible Steiner point $v$, the algorithm tries all possible lengths $j$ and $k$ to the first buffer on the path from $v$ to $t_i^1$ and respectively to $t_i^2$.

## 5   Extensions

In this section we describe how our approach to buffered global routing can be extended to handle pin assignment, polarity constraints imposed by the use of inverting buffers, buffer and wire sizing and prescribed delay upperbounds. The modifications required to handle these extensions involve only changes to the gadget graph described in Section 3.1, but not to the approximation algorithm in Figure 1.3 or to the randomized rounding scheme.

## 5.1 Pin Assignment

At the early stages of the design flow there is a significant degree of flexibility available for pin assignment, and therefore the ability to exploit this flexibility can have a major impact on the quality of resulting global routings. Consideration of pin assignment requires only two small changes in the construction of the gadget graph described in Section 3.1: (1) source vertices $s_i$ must now be connected by directed arcs to the $U$-th copies of *all* nodes representing enclosing tiles, and (2) copies $0, \ldots, U$ of *all* nodes representing enclosing tiles must be connected by directed arcs into the sink vertices $t_i$. Reading pin assignments from the paths selected by randomized rounding is trivial: we assign to each source (sink) an arbitrary pin in the tile visited first (last) by the selected path for the net.

The size of the gadget graph remains virtually unaffected by the pin assignment modification: for $k$ nets we only add $O(k)$ edges to the gadget graph under the realistic assumption that each pin can be assigned to at most $O(1)$ tiles. Therefore, the time required to find minimum-weight paths, and hence the overall runtime of the algorithm in Figure 1.3, does not increase even though the number of paths available for each net increases when considering pin assignment.

## 5.2 Polarity Constraints

The basic problem formulation in Section 2 considers only a non-inverting buffer type. In practice, inverting buffers are often preferred since they occupy a smaller area for the same driving strength. Although the use of inverting buffers introduces additional *polarity constraints*, which may require a larger number of buffers to be inserted, overall inverting buffers may lead to a better overall resource utilization. Algorithms for bounded capacitive load inverting (and non-inverting) buffer insertion have been recently discussed in [6]. The focus of [6] is on single net buffering, with arbitrary positions for the buffers. In our approach, the goal is to minimize the overall number of buffers required by the nets, and buffers can be inserted only in the available sites.

Consideration of polarity constraints require is achieved by modifying the basic gadget graph given in Section 3.1 as follows (see Figure 1.6). Each node of the basic gadget is replaced by an "even" and "odd" copy, i.e., $v^i$ is propagated into $v^i_{even}$ and $v^i_{odd}$. Tile-to-tile arcs are replaced by two arcs connecting copies with the same polarity, e.g., the arc $(u^i, v^{i-1})$ gives rise to $(u^i_{even}, v^{i-1}_{even})$ and $(u^i_{odd}, v^{i-1}_{odd})$. If a path uses such an arc, then it does not change polarity. Instead, each buffer arc changes polarity, i.e., $(v^i, v^U)$ gives rise to $(v^i_{even}, v^U_{odd})$ and $(v^i_{odd}, v^U_{even})$. The gadget also allows two inverting buffers to be inserted in the same tile for the purpose of meeting polarity constraints. This is achieved by providing bidirectional arcs connecting the $U$-th even and odd copies of a tile $v$, i.e., $(u^U_{even}, u^U_{odd})$ and $(v^U_{odd}, v^U_{even})$. Finally, source vertices $s_i$ are connected by directed arcs to the even $U$-th copy of enclosing tiles, and only copies of the desired polarity have arcs going into sink vertices $t_i$.

## 5.3 Buffer and wire sizing

Buffer and wire sizing are well-known techniques for timing optimization in the final stages of the design cycle [9]. However, early buffer and wire sizing can be equally effective for reducing congestion and/or wiring resources. In this section we show how to incorporate buffer and wire sizing during in our algorithmic framework for global buffered routing. The key enablers to these extensions are again appropriate modifications of the gadget graph.

The gadget for buffer sizing is illustrated in Figure 1.7(a) for two available buffer sizes, one with wireload upperbound $U = 4$ and one with wireload upperbound $U = 2$. The general construction entails using a number of copies of each tile vertex equal to the maximum buffer load upperbound $U$. For every buffer with wireload upperbound of $U' \leq U$, we insert buffer arcs $(v^i, v^{U'})$ for every $0 \leq i < U'$. Thus, the copy number of each vertex continues to capture the remaining wireload budget, which ensures the correctness of the construction.

Wire sizing (and a coarse form of layer assignment) can be handled by a different modification of the gadget graph (see Figure 1.7(b)). Assuming that per unit capacitances of the thinner wire widths are rounded to integer multiples of the "standard" per unit capacitance, the gadget models the use of thinner segments of wire by providing tile-to-tile arcs which decrease the tile copy index (i.e., remaining wireload budget) by more than one unit. For example, in Figure 1.7(b), solid arcs $(u^i, v^{i-1})$ and $(v^i, u^{i-1})$ correspond to standard width connections between tiles $u$ and $v$, while dashed arcs $(u^i, v^{i-2})$ and $(v^i, u^{i-2})$ correspond to "half-width" connections, i.e., connections using wire with double capacitive load per unit.

While our models for buffer and wire sizing are rather coarse (e.g., we truncate all buffer wireload upperbounds to integer multiples of the tile dimension, ignore variations in input capacitances of buffers and sinks, etc.), we consider them to be sufficiently accurate first-approximations for driving these optimizations during the early physical design stages.

## 5.4  Delay Constraints

In [3] we have proposed a method for enforcing given sink delay constraints based on charging wiresegment delays to buffer arcs in the gadget graph, and using a routine for computing minimum-weight delay constrained paths instead of Dijkstra's algorithm in the algorithm for approximating the fractional solution to ILP (1.1). Here we give a different method for handling sink delay constraints. The new method is similar in spirit to the constructions in previous sections, relying exclusively on a modification of the gadget graph. In general, our construction applies for any delay model, such as the Elmore delay model, for which (1) the delay of a buffered path is the sum of the delays of the path segments separated by the buffers, and (2) the delay of each segment depends only on segment length and buffer parameters. However, for the sake of efficiency, segment delays would have to be rounded to relatively coarse units.

Figure 1.8 shows the gadget construction for the case when delay is measured simply by the number of inserted buffers. The idea is again to replicate the basic gadget construction, this time a number of times equal to the maximum allowed net delay. Within each replica, tile-to-tile arcs decrease remaining wireload budget by one unit. In order to keep track of path delays, buffer arcs advance over a number of gadget replicas equal to the delay of the wiresegment ended by the respective buffer (this delay can be easily determined for each buffer arc since the tail of the arc fully identifies the length of the wiresegment). The construction is completed by connecting net sources to the vertices with maximum remaining wireload budget in the "0 delay" replica of the gadget graph, and adding arcs into the sinks from all vertices in replicas corresponding to delays smaller than the given delay upperbounds.

**Remark.** An interesting feature of the resulting gadget graph is that it is acyclic. Hence, we can now compute minimum-weight paths in the approximation algorithm in Figure 1.3 by computing the distances from the source via a topological traversal of the graph in $O(m+n)$ time instead of the $O(m+n\log n)$ time needed by Dijkstra's algorithm.

## 6  Practical enhancements

In this section we point to several enhancements which can improve the overall speed and accuracy of our algorithm.

**Uneven tile size.** Increasing the accuracy of our method can be achieved by decreasing the tile size. However, this results in significant increases in running time. Furthermore, when the tile size decreases beyond a certain point, the channel widths and the number of buffer sites per tile may become so small that the accuracy of the randomized rounding is greatly reduced. Ideally, the channel widths and buffer sites per tile should be approximately the same for all tiles. Indeed, if a tile is too crowded, then we can miss potential congestion violations, and if a tile is too sparse then the solution of LP relaxation cannot be rounded accurately. Our approach is to use *uneven* tile sizes in order to achieve evenly populated tiles. This approach can be implemented by choosing appropriate target values for channel width and buffer sites per tile and, starting with a coarse grid, recursively partitioning the overpopulated tiles into 4 equal sub-tiles until the target tile occupancy is reached.

**Window buffer constraints.** Our model assumes that the block design reserves room for buffer sites such that each block tile is assigned a specific number of buffer sites. Our approach can handle not only constraints on the number of buffer sites in *each tile* but additional constraints on the total number of buffers in a *set of tiles* (i.e., *windows*). For instance, these additional constraints may explicitly bound total number of buffers in a given block.

**Improved dual-update rules.** The algorithm in Figure 1.3 uses a multiplicative update rule for the dual variables: In each phase the dual variable corresponding to a set of edges $E'$ is multiplied by a factor of $(1+\varepsilon x)$, where $x$ is the ratio between the flow increase through $E'$ and the capacity of $E'$. This is not the only update rule that guarantees convergence. Another example is to update the dual corresponding to $E'$ by $e^{\varepsilon x}$. In [2] it is experimentally found that, for a similar multicommodity flow algorithm, this exponential update rule is more robust, i.e., guarantees convergence over a wider range of values for $\varepsilon$. Further improvements in runtime and solution quality can be achieved by computing an optimum update factor in each phase using Newton's method (see [2] for details).

## 7 Experimental results

In this section we report results for a 2-pin net implementation of our multicommodity flow based algorithm. All experiments with our algorithm were conducted on a 360 MHz SUN Ultra 60 workstation with 2 GB of memory, running under SunOS 5.7. The algorithm was coded in C and compiled using `g++` 3.2 with `-O4` optimization. Unless otherwise noted, the value of precision parameter $\varepsilon$ in the multicommodity flow algorithm was set to 0.3, and the number of iterations was limited to 64.

We tested the algorithm on the 10 circuits from [4], which are derived from testcases first used by [9]. We used the same circuit parameters as in the experiments for 2-pin nets of [4]; these parameters are summarized in Table 1.1. As in [4] and [9], we decomposed multipin nets into 2-pin nets by making direct connections from the source of a net to each of the net's sinks. Therefore, the numbers of 2-pin nets in Table 1.1 correspond to the numbers of sinks in [4]. We note that these numbers are slightly smaller than those reported in [9] since [4] retained only the nets requiring buffer insertion under the algorithm of [9]. We also note that the numbers of buffer sites used in our experiments are those used by Alpert et al. [4] in the experiments in which multi-pin nets were decomposed into 2-pin nets. These numbers were obtained directly from the authors, as they do not appear explicitly in [4] (the numbers of buffer sites given in Table 1 of [4] were used only in their experiments with un-decomposed multi-pin nets; see [5] for more details on these experiments). Finally, we note that although we use the same grid sizes as [4], there are some small differences in tile areas between Table 1.1 and [4]. These differences, which are probably due to the different procedures used in rounding tile dimensions, are unlikely to affect to a measurable degree the results of the compared algorithms.

Table 1.2 shows the results of the multicommodity flow algorithm (with pin assignment) when run with $D = \infty$, i.e., when the objective is to minimize the wire and buffer congestion only. The table shows that progressively better fractional solutions are obtained by the approximation algorithm. The results also show the tradeoff between congestion on one hand and wiring resources (number of buffers and wirelength) on the other hand.

Table 1.3 gives the results for wirelength minimization (i.e., using $\alpha = 0$ and $\beta = 1$) subject to wire and buffer congestion constraints ($\mu_0 = 1.0$ and $\nu_0 = 1.0$). In these experiments the multicommodity flow algorithm is run once per testcase (without binary search), with $D$ equal to the lower bound computed by routing each net optimally without taking into account capacity constraints. The multicommodity flow runtime includes randomized rounding (10000 trials, as described in Section 3.3). RABID runtime is for an RS6000/595 workstation with 1Gb of memory, as reported in [4].

The wirelength of the global routing obtained by our algorithm without pin assignment (MCF) is always within 1.03% of the lower bound. In contrast, the RABID heuristic of [4] exceeds the lower bound by 5.64 − 11.87%. To evaluate the effect of simultaneous pin assignment, we have added the possibility for each sink to be positioned not only in the given tile, but also in the 3-8 surrounding tiles (see Table 1 for the average number of tiles per pin of each testcase). Running our algorithm with pin assignment enabled (MCF+PA) further decreases wirelength by ≈ 10%, while being within at most 0.15% of the corresponding lower bound. We note that routing and pin assignment is performed by our algorithm in virtually the same time as routing alone.

Tables 1.4 gives results for the extension of the multicommodity flow algorithm to inverting buffer insertion, which is about twice slower than non-inverting buffer insertion due to the doubling in size of the gadget graph. Inverter insertion leads to a very small increase in the number of buffers (due to the need to satisfy polarity constraints) which is easily compensated by the smaller size of inverters. At the same time, inverter insertion requires virtually the same wirelength and often gives improved congestion (except for the xerox testcase).

Table 1.5 gives runtime scaling results for the extension of the multicommodity flow algorithm to buffered global routing with delay constraints. We note that the algorithm becomes faster for very tight delay constraints, since the number of nets that can meet delay constraints is only a fraction of the total number of nets. For moderately tight delay constraints almost all nets become routable, yet the runtime is comparable to that of the unconstrained version of the algorithm. For very lax delay constraints all nets become routable, and the runtime becomes significantly higher than that of the delay-oblivious version of the algorithm, by a factor roughly proportional to the increase in the size of the gadget graph, i.e., the maximum delay upperbound. However, large delay constraints are not very useful since they are satisfied almost in totality by using the unconstrained version of the algorithm.

## 8  Conclusions

In this paper we have presented the first provably good approach to buffered global routing with simultaneous timing- and congestion-driven buffered global route planning, pin and layer assignment, and wire/buffer sizing. The experimental results show that our method significantly outperforms approaches based on cascading individual optimizations such as the recent RABID algorithm of Alpert et al. [4]. Future work aims to incorporate in our implementation practical improvements such as the use of uneven sized tiles, window constraints on buffer usage (as opposed to tile constraints), and faster-converging dual-update rules.

## Acknowledgments

This work has been supported in part by Cadence Design Systems, Inc., the MARCO Gigascale Silicon Research Center, and NSF Grant CCR-9988331. Preliminary versions of these results have appeared in [3, 15]. The authors wish to thank Charles Alpert, Jason Cong, and Jiang Hu for kindly providing us with the testcases used in [9] and [4].

Table 1.1: Circuit parameters.

| Circuit | # 2-Pin Nets | Grid size | Tile area | $w(e)$ | Avg. tiles per pin | U | #Buffer sites |
|---:|---:|:---:|---:|---:|---:|---:|---:|
| a9c3    | 1526 | 30 x 30 | 1.09 | 52  | 4.9 | 6 | 32780 |
| ac3     | 409  | 30 x 30 | 0.49 | 26  | 5.0 | 7 | 8550  |
| ami33   | 324  | 33 x 30 | 0.46 | 32  | 5.0 | 6 | 17750 |
| ami49   | 493  | 30 x 30 | 0.68 | 14  | 4.8 | 6 | 11450 |
| apte    | 141  | 30 x 33 | 0.36 | 13  | 5.0 | 7 | 4200  |
| hc7     | 1318 | 30 x 30 | 1.04 | 28  | 4.8 | 6 | 17780 |
| hp      | 187  | 30 x 30 | 0.42 | 12  | 5.0 | 7 | 2350  |
| playout | 1663 | 33 x 30 | 0.78 | 120 | 4.8 | 7 | 37550 |
| xc5     | 2149 | 30 x 30 | 0.58 | 50  | 5.0 | 7 | 19150 |
| xerox   | 390  | 30 x 30 | 0.38 | 40  | 5.0 | 7 | 7000  |

Table 1.2: Congestion minimization results ($D = \infty$) for the multicommodity flow algorithm with $\varepsilon = 0.3$.

| Circuit | Phase# | Wire Congest | Buffer Congest | #Buffers | Wlen | CPU sec. |
|---|---|---|---|---|---|---|
| a9c3 | 1 | 0.75 | 0.80 | 3351 | 26057 | 12.0 |
|  | 4 | 0.59 | 0.43 | 3356 | 26123 | 47.5 |
|  | 16 | 0.51 | 0.23 | 3402 | 26595 | 188.8 |
|  | 64 | 0.46 | 0.18 | 3505 | 27328 | 730.7 |
|  | **64+ROUND** | **0.62** | **0.30** | **3625** | **27980** | **785** |
| ac3 | 1 | 0.77 | 1.00 | 796 | 4998 | 3.0 |
|  | 4 | 0.62 | 0.53 | 797 | 5008 | 12.2 |
|  | 16 | 0.40 | 0.27 | 803 | 5072 | 48.9 |
|  | 64 | 0.28 | 0.18 | 826 | 5211 | 192.3 |
|  | **64+ROUND** | **0.46** | **0.50** | **827** | **5251** | **213** |
| ami33 | 1 | 0.66 | 0.67 | 909 | 4466 | 2.6 |
|  | 4 | 0.55 | 0.36 | 908 | 4476 | 10.6 |
|  | 16 | 0.47 | 0.20 | 910 | 4515 | 42.2 |
|  | 64 | 0.40 | 0.14 | 930 | 4618 | 163.0 |
|  | **64+ROUND** | **0.56** | **0.31** | **956** | **4698** | **181** |
| ami49 | 1 | 1.36 | 0.90 | 948 | 6045 | 3.0 |
|  | 4 | 1.00 | 0.46 | 958 | 6083 | 12.3 |
|  | 16 | 0.74 | 0.29 | 1040 | 6509 | 52.1 |
|  | 64 | 0.66 | 0.21 | 1205 | 7278 | 211.3 |
|  | **64+ROUND** | **1.00** | **0.56** | **1308** | **7751** | **234** |
| apte | 1 | 1.08 | 1.00 | 328 | 1668 | 1.2 |
|  | 4 | 0.87 | 0.57 | 327 | 1677 | 5.0 |
|  | 16 | 0.53 | 0.30 | 336 | 1725 | 21.1 |
|  | 64 | 0.44 | 0.17 | 359 | 1836 | 86.8 |
|  | **64+ROUND** | **1.00** | **1.00** | **360** | **1841** | **98** |
| hc7 | 1 | 1.00 | 1.19 | 2203 | 17670 | 8.0 |
|  | 4 | 0.79 | 0.61 | 2206 | 17738 | 32.5 |
|  | 16 | 0.69 | 0.31 | 2301 | 18481 | 132.9 |
|  | 64 | 0.62 | 0.23 | 2498 | 19660 | 516.9 |
|  | **64+ROUND** | **0.89** | **0.50** | **2675** | **20584** | **562** |
| hp | 1 | 0.92 | 1.67 | 334 | 1952 | 1.3 |
|  | 4 | 0.71 | 0.85 | 330 | 1961 | 5.2 |
|  | 16 | 0.46 | 0.45 | 334 | 2003 | 21.8 |
|  | 64 | 0.33 | 0.29 | 355 | 2119 | 89.8 |
|  | **64+ROUND** | **0.58** | **1.00** | **362** | **2147** | **101** |
| playout | 1 | 0.64 | 0.98 | 2890 | 23155 | 14.9 |
|  | 4 | 0.52 | 0.42 | 2892 | 23199 | 60.6 |
|  | 16 | 0.40 | 0.24 | 2922 | 23582 | 257.5 |
|  | 64 | 0.33 | 0.17 | 3238 | 25809 | 1118.8 |
|  | **64+ROUND** | **0.36** | **0.28** | **3467** | **27281** | **1176** |
| xc5 | 1 | 1.14 | 1.31 | 3187 | 22314 | 17.6 |
|  | 4 | 0.98 | 0.66 | 3202 | 22492 | 70.9 |
|  | 16 | 0.74 | 0.37 | 3277 | 23231 | 288.2 |
|  | 64 | 0.66 | 0.31 | 3570 | 24872 | 1134.7 |
|  | **64+ROUND** | **0.88** | **0.57** | **3895** | **26305** | **1216** |
| xerox | 1 | 0.93 | 1.42 | 659 | 3662 | 2.8 |
|  | 4 | 0.72 | 0.77 | 660 | 3698 | 11.9 |
|  | 16 | 0.45 | 0.40 | 684 | 3858 | 54.0 |
|  | 64 | 0.32 | 0.21 | 753 | 4174 | 237.6 |
|  | **64+ROUND** | **0.47** | **0.67** | **779** | **4295** | **257** |

Table 1.3: Wirelength minimization ($\alpha = 0$ and $\beta = 1$) subject to wire and buffer congestion constraints ($\mu_0 = 1.0$ and $\nu_0 = 1.0$). RABID is the algorithm of [4], MCF is the multicommodity flow algorithm without pin assignment capability, and MCF+PA is the multicommodity flow algorithm with pin assignment enabled. Both MCF and MCF+PA were run with $\varepsilon = 0.3$.

| Circuit | Algorithm | Wlen | %LB gap | #buffers | %LB gap | Wire Congest | Buffer Congest | CPU sec. |
|---|---|---|---|---|---|---|---|---|
| a9c3 | RABID | 30723 | 5.64 | 4225 | 11.95 | 0.60 | 0.44 | 502 |
|  | MCF+ROUND | 29082 | 0.00 | 3800 | 0.69 | 0.67 | 0.31 | 775 |
|  | MCF+PA+ROUND | 26057 | 0.00 | 3378 | 0.81 | 0.62 | 0.30 | 779 |
| ac3 | RABID | 5954 | 7.67 | 1037 | 15.74 | 0.58 | 0.33 | 208 |
|  | MCF+ROUND | 5530 | 0.00 | 905 | 1.00 | 0.77 | 0.67 | 204 |
|  | MCF+PA+ROUND | 4993 | 0.00 | 803 | 1.13 | 0.69 | 0.67 | 204 |
| ami33 | RABID | 5232 | 6.93 | 1150 | 14.20 | 0.69 | 0.44 | 138 |
|  | MCF+ROUND | 4893 | 0.00 | 1014 | 0.70 | 0.75 | 0.33 | 177 |
|  | MCF+PA+ROUND | 4464 | 0.00 | 916 | 0.88 | 0.53 | 0.50 | 177 |
| ami49 | RABID | 7592 | 11.87 | 1339 | 21.51 | 0.93 | 0.36 | 167 |
|  | MCF+ROUND | 6792 | 0.07 | 1133 | 2.81 | 1.00 | 0.60 | 227 |
|  | MCF+PA+ROUND | 6041 | 0.01 | 989 | 4.66 | 1.00 | 0.44 | 218 |
| apte | RABID | 2010 | 10.78 | 417 | 18.47 | 1.00 | 0.33 | 95 |
|  | MCF+ROUND | 1833 | 1.03 | 377 | 7.10 | 1.00 | 1.00 | 88 |
|  | MCF+PA+ROUND | 1663 | 0.15 | 331 | 4.75 | 1.00 | 1.00 | 87 |
| hc7 | RABID | 21523 | 7.54 | 2983 | 17.44 | 0.82 | 0.35 | 386 |
|  | MCF+ROUND | 20024 | 0.05 | 2591 | 2.01 | 0.96 | 0.47 | 551 |
|  | MCF+PA+ROUND | 17660 | 0.00 | 2214 | 0.68 | 0.86 | 0.47 | 543 |
| hp | RABID | 2403 | 11.12 | 450 | 20.97 | 0.83 | 0.28 | 67 |
|  | MCF+ROUND | 2165 | 0.13 | 404 | 8.60 | 1.00 | 1.00 | 95 |
|  | MCF+PA+ROUND | 1945 | 0.00 | 345 | 6.81 | 0.92 | 1.00 | 94 |
| playout | RABID | 27601 | 6.38 | 3840 | 15.04 | 0.45 | 0.64 | 813 |
|  | MCF+ROUND | 25946 | 0.00 | 3429 | 2.73 | 0.53 | 0.34 | 1002 |
|  | MCF+PA+ROUND | 23138 | 0.00 | 3011 | 4.37 | 0.40 | 0.32 | 995 |
| xc5 | RABID | 27060 | 8.35 | 4410 | 23.25 | 0.84 | 0.81 | 694 |
|  | MCF+ROUND | 25151 | 0.71 | 3843 | 7.41 | 0.98 | 0.62 | 1162 |
|  | MCF+PA+ROUND | 22265 | 0.05 | 3341 | 4.90 | 1.00 | 0.65 | 1175 |
| xerox | RABID | 4541 | 11.48 | 957 | 30.56 | 0.93 | 0.57 | 167 |
|  | MCF+ROUND | 4078 | 0.12 | 805 | 9.82 | 1.00 | 1.00 | 212 |
|  | MCF+PA+ROUND | 3658 | 0.00 | 692 | 6.30 | 0.88 | 0.67 | 208 |

Table 1.4: Wirelength minimization results for non-inverting vs. inverting buffer insertion. The number of buffer sites was assumed to be the same in both experiments.

| Testcase | Non-inverting buffers | | | | | Inverting buffers | | | | |
|---|---|---|---|---|---|---|---|---|---|---|
|  | Wlen | #buffers | W-Congest | B-Congest | CPU | Wlen | #buffers | W-Congest | B-Congest | CPU |
| a9c3 | 29082 | 3800 | 0.67 | 0.31 | 775 | 29082 | 4540 | 0.60 | 0.41 | 1470 |
| ac3 | 5530 | 905 | 0.77 | 0.67 | 204 | 5530 | 1095 | 0.69 | 0.50 | 417 |
| ami33 | 4893 | 1014 | 0.75 | 0.33 | 177 | 4893 | 1186 | 0.62 | 0.29 | 359 |
| ami49 | 6792 | 1133 | 1.00 | 0.60 | 227 | 6790 | 1417 | 1.00 | 1.00 | 449 |
| apte | 1833 | 377 | 1.00 | 1.00 | 88 | 1833 | 441 | 1.00 | 1.00 | 185 |
| hc7 | 20024 | 2591 | 0.96 | 0.47 | 551 | 20024 | 3358 | 0.89 | 0.50 | 1030 |
| hp | 2165 | 404 | 1.00 | 1.00 | 95 | 2164 | 495 | 1.00 | 1.00 | 201 |
| playout | 25946 | 3429 | 0.53 | 0.34 | 1002 | 25946 | 4235 | 0.53 | 0.32 | 1982 |
| xc5 | 25151 | 3843 | 0.98 | 0.62 | 1162 | 25222 | 4799 | 0.96 | 1.00 | 2285 |
| xerox | 4078 | 805 | 1.00 | 1.00 | 212 | 4155 | 1050 | 1.18 | 1.00 | 520 |

Table 1.5: Runtime scaling for the timing-driven version of the MCF algorithm (delay measured by number of inserted buffers).

| Testcase | Delay bound = 1 | | Delay bound = 2 | | Delay bound = 4 | | Delay bound = 8 | | No delay bound | |
|---|---|---|---|---|---|---|---|---|---|---|
| | #nets | CPU | #nets | CPU | #nets | CPU | #nets | CPU | #nets | CPU |
| a9c3 | 455 | 77 | 820 | 361 | 1361 | 2178 | 1526 | 7667 | 1526 | 775 |
| ac3 | 152 | 29 | 249 | 122 | 374 | 666 | 409 | 2270 | 409 | 204 |
| ami33 | 63 | 13 | 125 | 50 | 260 | 413 | 323 | 1761 | 324 | 177 |
| ami49 | 177 | 25 | 298 | 113 | 442 | 598 | 493 | 2161 | 493 | 227 |
| apte | 49 | 12 | 67 | 33 | 126 | 255 | 141 | 968 | 141 | 88 |
| hc7 | 569 | 70 | 873 | 305 | 1231 | 1584 | 1318 | 5217 | 1318 | 551 |
| hp | 76 | 15 | 124 | 63 | 174 | 330 | 187 | 1083 | 187 | 95 |
| playout | 657 | 124 | 1095 | 651 | 1575 | 3506 | 1663 | 10979 | 1663 | 1002 |
| xc5 | 1072 | 192 | 1429 | 748 | 2100 | 4158 | 2149 | 12351 | 2149 | 1162 |
| xerox | 163 | 29 | 282 | 201 | 360 | 752 | 390 | 2308 | 390 | 212 |

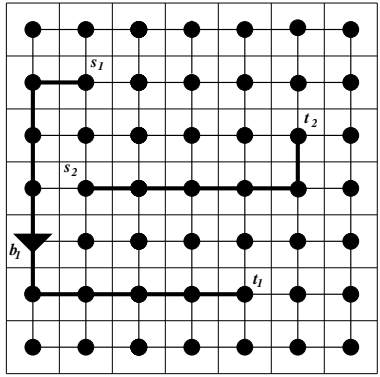

Figure 1.1: Tile graph with two 2-pin nets.

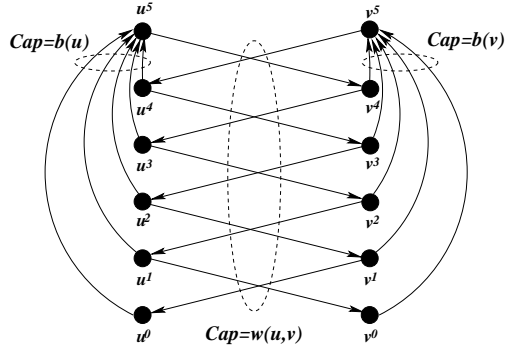

Figure 1.2: The basic gadget replacing edge $(u,v)$ of the tile graph for buffer wireload upperbound $U = 5$.

(1) Set $y_v := \frac{\delta}{\mu_0 b(v)} \ \forall v \in V(G), \quad z_e := \frac{\delta}{\nu_0 w(e)} \ \forall e \in E(G), \quad u := \frac{\delta}{D}$
(2) Set $x_p := 0 \ \forall p \in \mathcal{P}$
(3) Set $r = 0$ and $p_i := \emptyset$ for $i = 1, ..., k$.
(4) While $\mu_0 \sum_{v \in V(G)} b(v) y_v + \nu_0 \sum_{(u,v) \in E(G)} w(u,v) z_{u,v} + Du < 1$ do:
(5) **begin**
(6) r := r+1.
(7) For $i := 1$ to $k$, do
(8) **begin**
(9) If $p_i = \emptyset$ or $\sum_{v \in V(G)} |p_i \cap E_v|(y_v + \alpha u) + \sum_{(u,v) \in E(G)} |p_i \cap E_{u,v}|(z_{u,v} + \beta u) > (1+\gamma\varepsilon) l_i$ then
(10) **begin**
(11) Find a path $p_i \in \mathcal{P}_i$ minimizing $l_i := \sum_{v \in V(G)} |p_i \cap E_v|(y_v + \alpha u) + \sum_{(u,v) \in E(G)} |p_i \cap E_{u,v}|(z_{u,v} + \beta u)$
(12) **end**
(13) Set $x_{p_i} := x_{p_i} + 1$
(14) Set $y_v := y_v \left(1 + \varepsilon \frac{|p_i \cap E_v|}{\mu_0 b(v)}\right) \forall v \in V(G), \quad z_e := z_e \left(1 + \varepsilon \frac{|p_i \cap E_{u,v}|}{\nu_0 w(u,v)}\right) \forall (u,v) \in E(G)$
$$u := u \left(1 + \varepsilon \frac{\alpha \sum_{v \in V(G)} |p_i \cap E_v| + \beta \sum_{(u,v) \in E(G)} |p_i \cap E_{u,v}|}{D}\right)$$
(15) **end**
(16) **end**
(17) Output $(x_p/r)_{p \in \mathcal{P}}$

Figure 1.3: Algorithm for approximately solving LP (1.2).

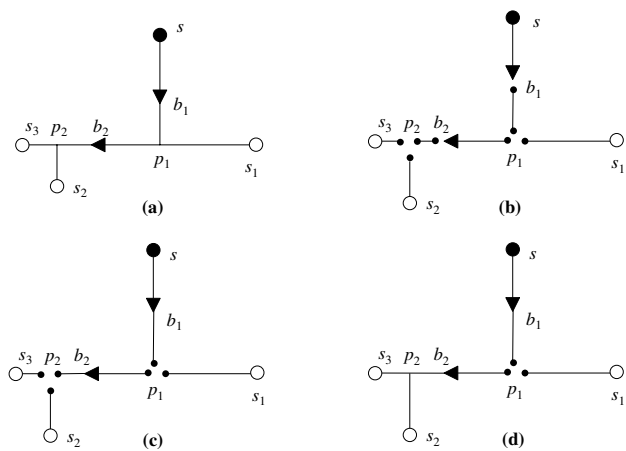

Figure 1.4: (a) Minimum buffered routing $T$ of a 4-pin net with source $s$ and sinks $s_1, s_2, s_3$ produced by the algorithm in [6]. $T$ has 2 buffers $b_1$ and $b_2$ and two Steiner points $p_1$ and $p_2$. (b) Decomposition of $T$ using fixed branching and fixed buffering. There are 7 resulting 2-pin nets: $(s, b_1), (b_1, p_1), (p_1, s_1), (p_1, b_2), (b_2, p_2), (p_2, s_2)$, and $(p_3, s_3)$. (c) Decomposition of $T$ using fixed branching and flexible buffering. There are 5 resulting 2-pin nets: $(s, p_1), (p_1, s_1), (p_1, p_2), (p_2, s_2)$, and $(p_3, s_3)$. (d) Decomposition of $T$ using half-flexible branching and flexible buffering. The Steiner point $p_1$ is fixed and, therefore, split between 3 nets, while the Steiner point $p_2$ is flexible. There are 2 resulting 2-pin nets: $(s, p_1), (p_1, s_1)$ and one 3-pin net $(p_1, s_2, s_3)$.

```
(1)  Set w* := ∞
(2)  For all v ∈ V do      // try all possible Steiner points
(3)  begin
(4)    For j := 0 to U
(5)    begin
(6)      Find a shortest v^{U-j} − t_i^1–path P_1 in H
(7)      For k := 0 to U − j
(8)      begin
(9)        Find a shortest v^{U-k} − t_i^2–path P_2 in H
(10)       Find a shortest s_i^0 − v^{U-j-k}–path P_0 in H
(11)       If w(P_0) + w(P_1) + w(P_2) ≤ w* then
(12)         Set w* := w(P_0) + w(P_1) + w(P_2); T* := P_0 ∪ P_1 ∪ P_2
(13)     end
(14)   end
(15) end
(16) return T*
```

Figure 1.5: Algorithm for finding minimum weight buffered routings for 3-pin nets.

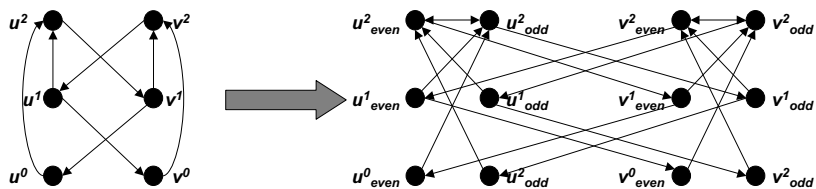

Figure 1.6: Gadget for polarity constraints with buffer load upperbound $U = 2$.

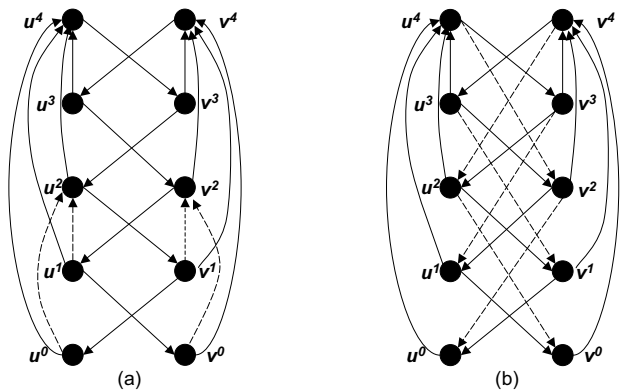

Figure 1.7: (a) Gadget for buffer sizing with two available buffer sizes, one with wireload upperbound $U = 4$ and one with wireload upperbound $U = 2$. Solid arcs $(u^i, u^4)$, respectively $(v^i, v^4)$, correspond to the insertion of a buffer capable of driving 4 units of wire, while dashed arcs $(u^i, u^2)$ and $(v^i, v^2)$ correspond to the insertion of a smaller buffer capable of driving 2 units of wire. (b) Gadget for wire sizing with two available wire widths, standard width and "half" width (i.e., wire with double per unit capacitive load). The solid arcs $(u^i, v^{i-1})$ and $(v^i, u^{i-1})$ correspond to standard width connections between tiles $u$ and $v$, while dashed arcs $(u^i, v^{i-2})$ and $(v^i, u^{i-2})$ correspond to half-width connections.

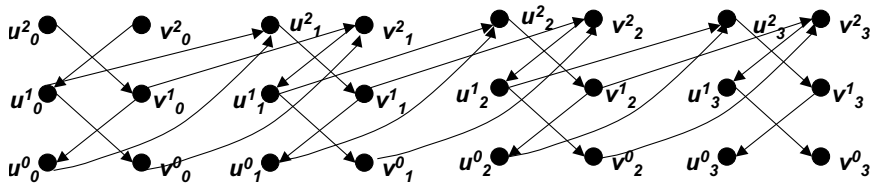

Figure 1.8: Gadget for enforcing delay constraints when the delay is measured by the number of buffers inserted between source and sink. The basic gadget is replicated a number of times equal to the maximum allowed net delay (3 in this example). Tile-to-tile arcs decrease remaining wireload budget within a gadget replica, while buffer arcs advance from one replica to the next.